# Semi-automatic System for Title Construction

Swagata Duari[0000-0002-9057-0140] and Vasudha Bhatnagar[0000-0002-9706-9340]

Department of Computer Science, University of Delhi, New Delhi, India - 110007
sduari@cs.du.ac.in

**Abstract.** In this paper, we propose a semi-automatic system for title construction from scientific abstracts. The system extracts and *recommends* impactful words from the text, which the author can creatively use to construct an appropriate title for the manuscript. The work is based on the hypothesis that keywords are good candidates for title construction.

We extract important words from the document by inducing a supervised keyword extraction model. The model is trained on novel features extracted from graph-of-text representation of the document. We empirically show that these graph-based features are capable of discriminating keywords from non-keywords. We further establish empirically that the proposed approach can be applied to any text irrespective of the training domain and corpus. We evaluate the proposed system by computing the overlap between extracted keywords and the list of title-words for documents, and we observe a macro-averaged precision of 82%.

## 1 Introduction

The *title* of a scientific research article plays an important role in the process of literature review. It is the most important piece of information that assists the reader in sifting through vast amount of text in a repository. The title of a document conveys the central idea expressed in a document by establishing the premise of discussion contained in the text, and provides a clear yet simple one-line summary of the document content. Deciding on a title for scientific write-up or blog articles has always been a task of immense importance, because the reader often decides on the relevance of that document to his/her query just by observing the title. Writers often go through several iterations in order to decide upon the most satisfactory title for their article.

Studies have shown that the title of a scientific paper can influence the number of reads and citations for that article [10, 15, 21]. Paiva et al. reported that articles with short and result-oriented titles are more likely to be cited than those with long and method-describing titles [21]. Automatic generation of full-fledged title for scientific write-up is a complex process that requires natural language generation, which is still immature. We propose a semi-automatic method for *constructing* titles from scientific articles by identifying impactful words appearing in the abstracts. We hypothesize that

keywords express the crux of the document and are therefore likely to be a part of the document title. Thus, we propose an application of automatic keyword extraction where extracted keywords are recommended to the author, which can be used for title construction after suitable transformation and by including other glue words. It is noteworthy that our work is different from automatic title generation, where a full-fledged title is automatically generated for the document. Instead, the proposed system aids in 'constructing' the title by automatically extracting the important words from the text and suggesting them to the author.

*Keyword Extraction* (henceforth, KE) is a classic data mining problem that addresses the task of automatically extracting special words from text documents to present a compact and precise representation of the document content. Typically embedded in document text, keywords for research articles not only convey the topics that the document covers, but are also used by search engines and document databases to efficiently locate information. We propose to identify keywords for title construction using an automatic keyword extraction method. The novelty of our approach lies in the fact that we design a generic supervised keyword extraction model that can be applied on any text without considering its domain or corpora. We aim to achieve the goal by exploiting the advantages of graph-based keyword extraction methods. Specifically, our contributions are as given below.

i. We demonstrate that the properties of the words extracted from graph representation of text are effective features to discriminate between keywords and non-keywords.

ii. We note that simple classifiers perform well enough for the task of keyword extraction. Complex algorithms, such as deep learning, not necessarily yield better performance considering the small training sets and the training time.

iii. We show that the extracted keywords appear $\approx 80\%$ times in the title of scientific articles in our dataset.

The paper is organized as follows. We discuss works related to our research in Section 2, followed by methodology of the proposed algorithm in Section 3. Section 4 covers experimental setup, dataset details, objectives of each experiment, and preliminary results and discussions. Finally, Section 5 concludes our paper.

## 2 Related Works

In this section, we discuss works related to automatic title generation and automatic keyword extraction, which are relevant to our study.

### 2.1 Automatic Title Generation

Various studies have been performed on automatic title generation from both spoken [5, 11, 12] and written text [13, 14, 22, 25, 27]. These works aim at converting the document into a 'title representation' by using either statistical, probabilistic, or machine learning methods. Kennedy et al. used an EM-based Bayesian statistical machine-translation method to identify title-word and document-word pairs that are most likely to constitute the document title [14]. Jin et al. proposed a probabilistic approach for automatic title generation that takes into account title word ordering [13].

Automatic title generation has also been viewed as an automatic summarization task by some researchers. Tseng et al. applied the task of automatic title generation to document clustering in order to identify generic labels for better cluster interpretation [27]. In a recent study, Shao et al. used a dependency tree based method to identify and filter candidate sentences that are likely to be titles [25]. In a similar fashion, Putra et al. used adaptive KNN to produce a single-line summary from article abstracts and argued that rhetorical categories (research goal, method, and not relevant) of sentences in the document have potential to boost the title generation process.

### 2.2 Automatic Keyword Extraction

Automatic KE methods fall under two categories - supervised and unsupervised [1, 3]. Supervised methods treat keyword extraction problem as a binary classification task ('keyword' and 'non-keyword' classes), whereas unsupervised methods use statistical or graph-theoretic properties to rank candidate words.

The primary task in any supervised KE methods is to construct the feature set. Identifying good quality features that effectively discriminate keywords from non-keywords is a challenging task. Some of the popular features that are used in literature are tf-idf, POS tags, n-gram features, etc. [4, 9, 20, 26]. Apart from these, topical information [30], linguistic knowledge [9], structural features of the document [17], knowledge about domain and collection [4, 20], expert knowledge [8], and external sources like Wikipedia links [18] are used to enrich the feature set. Moreover, the objective of supervised KE methods is to identify potential key-*phrases*, and not key-*words*. We, however, slightly differ from rest of the state-of-the-art supervised KE methods and focus on identifying keywords instead of phrases.

In general, supervised approaches for keyword extraction report better results compared to unsupervised counterparts. Unsupervised KE techniques largely comprise graph-based methods, which transform the text into a graph and use graph-theoretic properties to rank keywords. Local node properties like PageRank [19], PageRank along with position of the word in text [7], degree centrality [16], coreness [23], etc. have been studied extensively in the past. Unlike supervised methods, the primary advantage of unsupervised methods is that they are independent of the domain or corpus of the document.

## 3 Methodology

In this paper, we propose a semi-automatic system for title construction. The system works in two phases. In the first phase, the system automatically extracts stemmed keywords from the text document and presents them to the author. These words are the candidates for title construction. In the second phase, which requires manual intervention, the author can creatively weave the title by suitably transforming the stemmed candidates and using glue words.

We design a supervised keyword extractor to implement the first phase by exploiting graph-theoretic properties of candidate keywords. We hypothesize that certain node properties are capable of distinguishing keywords and non-keywords, and accordingly transform the document to a graph-of-text representation. The proposed algorithm comprises of the following steps.

i.  Prepare the training set as follows.
    a)  Select candidate keywords from each document, and construct the corresponding graph-of-text (Section 3.1)
    b)  Extract select node properties from each graph-of-text and assign label to each candidate keyword based on the available gold-standard keywords list (Section 3.2).
    c)  Balance the training set, if required.
ii. Train a predictive model using the prepared training set and use it to predict keywords for target document (Section 3.3).
iii. Recommend top-$k$ extracted keywords as candidates for title construction.

The usage of the proposed system is summarized in Figure 1. To construct the title, the text is converted to a graph. Node properties of each word are extracted and are supplied to a pre-trained model, which outputs the probability of each word being a 'keyword'. Top-$k$ keywords suggested to the user can be used for title construction ($k$ is user specified).

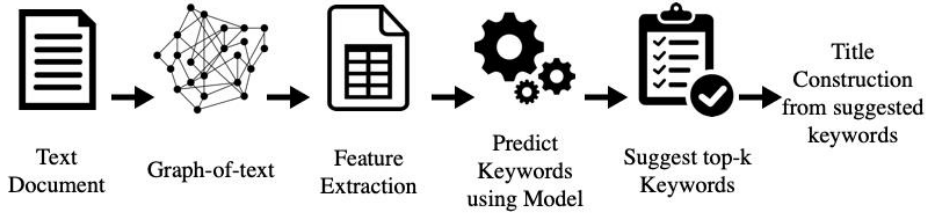

Text Document — Graph-of-text — Feature Extraction — Predict Keywords using Model — Suggest top-k Keywords — Title Construction from suggested keywords

**Fig. 1.** Diagrammatic representation of the semi-automatic system for title construction.

### 3.1 Candidate Selection and Modeling Text as Graphs

We follow the well-established convention to retain nouns and adjectives from the text as candidate keywords [19, 7, 23]. The text is then transformed to a graph representation, where the candidate keywords constitute the set of nodes and the set of links are defined based on a co-occurrence relation. Following Duari et al., we use a parameter-free approach for creating context aware graphs (CAG) [6], where links between nodes are forged if they co-occur within two consecutive sentences. The output graph is undirected, and is weighted by the number of times the adjacent nodes (words) co-occur in the original text. We exclude isolated nodes from computation. Please note that short texts (1-3 sentences) result into highly dense graphs, which are often complete graphs. Graph density decreases with increase in the number of sentences. Figure 2 shows CAG graph of a short example text. Edge width in the graph is proportional to corresponding edge weight.

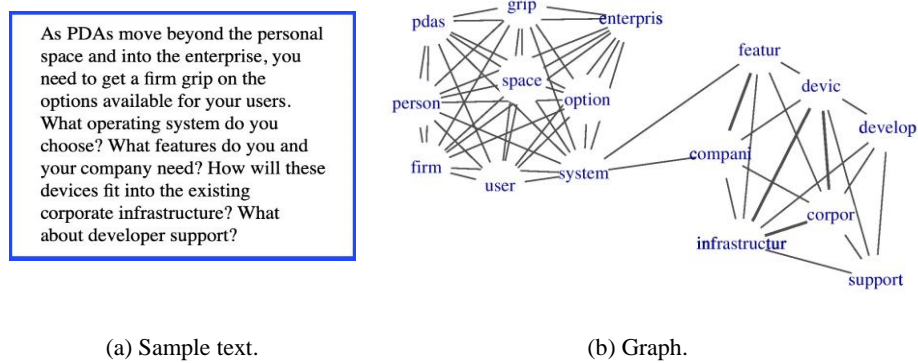

(a) Sample text.                                        (b) Graph.

**Fig. 2.** Graph created from sample text in Figure 2a (document id 1450 from Hulth2003 dataset).

### 3.2 Training set construction

Node centrality measures quantify the importance of a node in the graph, and are well studied in graph theory. Centrality is a local node property that estimates a node's embeddedness in the network. Intuitively, nodes with high centrality value are more important in the network. We consider Degree Centrality, Eigenvector Centrality (Prestige), and PageRank [2] as features. Additionally, we consider an extension of PageRank called PositionRank [7], a graph degeneracy method called Coreness [24], and Clustering Coefficient as features. For each node in the CAG graph, we compute the six properties to create the feature set.

We denote a weighted, undirected graph by $G = (V, E, W)$, with $V$ as the set of vertices, $E \in V \times V$ as the set of edges, and $W$ as the corresponding weighted adjacency matrix. Node properties used as features are briefly described below.

**Degree centrality** of a vertex measures its embedded-ness at local level. For a weighted graph, it is computed as $deg(v_i) = \sum_j w_{ij}$ [28].

**Prestige** or *Eigenvector centrality* of vertex $v_i$ quantifies its embedded-ness in the graph while (recursively) taking into account the prestige of its neighbors. Starting with initial prestige vector $\boldsymbol{p_0}$ where all nodes (words) are assigned equal *prestige*, $\boldsymbol{p_k}$ is computed recursively as follows till convergence is achieved [28].

$$\boldsymbol{p_k} = W^T \boldsymbol{p_{k-1}} = (W^k)^T \boldsymbol{p_0}$$

According to this computation, a well-connected word attains more importance if it is connected to other well-connected words.

**PageRank** computes prestige in the context of Web search with an additional component called *random jump*. In case of text documents, this component relates to the concept of text cohesion [19]. We adopt the computation of word score ($WS$) from TextRank algorithm [19], with $d = 0.85$ as the probability of random jump.

$$WS(v_i) = (1 - d) + d * \sum_{v_j \in N_i} \left( \frac{w_{ji}}{\sum_{v_k \in N_j} w_{jk}} WS(v_j) \right)$$

**PositionRank** is an extension of PageRank that favors words occurring at the beginning of the document as keywords [7]. Node $v_i \in V$ is assigned a weight $p_i$ based on its positional information by taking the inverse of the sum of its positions of occurrences in the text. Subsequently, PageRank computation is performed on the weighted nodes of the graph to yield PositionRank scores for the candidate words. Mathematically, the PositionRank score of a node $v_i$ is computed as follows.

$$S(v_i) = (1 - \alpha) . \tilde{p}_i + \alpha . \sum_{v_j \in N_i} \left( \frac{w_{ji}}{O(v_j)} S(v_j) \right)$$

Here, $\alpha$ is set to 0.85, $\tilde{p}_i = \frac{p_i}{\sum_{j=1}^{|V|} p_j}$ is the normalized positional weight of $v_i$, $N_i$ is the neighborhood of node $v_i$, $w_{ji}$ is the weight of edge $e_{ji}$, and $O(v_j) = \sum_{v_k \in N_j} w_{jk}$.

**Coreness** is a graph degeneracy property that decomposes graph $G$ into a set of maximal connected subgraphs $G_k$ ($k$ denotes the core), such that nodes in $G_k$ have degree at least $k$ within the subgraph and $G_k \subseteq G_{k-1}$ [24]. Coreness of a node is the highest core to which it belongs. Rousseau et al. [23] presume that words in the main (highest) core of the graph are keywords due to their dense connections. Though our findings differ, we are convinced that keywords tend to lie in higher cores. Hence, we choose to include coreness as a discriminating property.

**Clustering Coefficient** of a node indicates edge density in its neighborhood. Clustering coefficient for node $v_i$ is computed as the ratio of actual number of edges in the subgraph induced by $v_i$ (excluding itself) to the total number of possible edges in that subgraph [28]. Mathematically, for an undirected graph $G$, clustering coefficient of node $v_i \in G$ is computed as below.

$$CC(v_i) = \frac{2 \left| e_{jk} : v_j, v_k \in N_i, e_{jk} \in E \right|}{n_i(n_i - 1)}$$

Here, $n_i$ is the number of nodes in $N_i$, i.e., the subgraph induced by $v_i$. We speculate that nodes (words) with low clustering coefficient connect diverse contents together, and thus are likely to be important words.

All properties, except Clustering Coefficient, have been studied by state-of the-art unsupervised graph-based keyword extraction methods. To the best of authors' knowledge, complex interplay of these properties has not been explored for discriminating between keywords and non-keywords.

**Assigning labels:** For each document, we consult the corresponding gold-standard keywords list and assign the label as 'positive' or 'negative' to the candidate words (nodes) depending on whether they are listed as a gold-standard keyword (as unigram) or not. The labels along with the feature set constitute the training set for our KE algorithm.

### 3.3 Model Training

We prepare three training sets using the steps described in Section 3.2 for the KDD, WWW, and Hulth2003 datasets (dataset details are given in Table 1). Training set for Hulth2003 dataset is relatively balanced. However, training sets for KDD and WWW datasets are imbalanced in nature. This is because on average, each document from KDD and WWW datasets is assigned $\approx$ 10 gold-standard keywords (unigrams) out of $\approx$ 100-200 words (columns $N_{avg}$ and $L_{avg}$ of Table 1, respectively). Since imbalanced dataset does not yield robust predictive model, we balance both these training sets by over-sampling the 'positive' class using Weka implementation of SMOTE filter[1]. Using the five training sets (two imbalanced and three balanced) as individual training sets, we train the predictive models.

Several classification algorithms have been explored in literature, including CRF and SVM [29], Bagged decision tree [18], Naïve Bayes [4, 26], gradient boosted decision trees [26], etc. However, we decided to use two classical algorithms - Naïve Bayes (NB) and Logistic Regression (LR) - because of their simplicity and fast execution time. Using these two algorithms and the five training sets, we induce ten (10) predictive models - five using NB and five using LR for each training set. Our empirical results validate that classical algorithms perform adequately for our experiments. We present the cross-validation and test results in Section 4.2.

## 4 Empirical Evaluation

In this section, we present our experimental setup and empirical results. We also discuss our findings, and empirically establish our claims.

### 4.1 Experimental Setup and Objectives

The proposed framework is implemented using R (version 3.4.0) and relevant packages[2] (`igraph`, `tm`, `openNLP`, `RWeka`, `caret` and `pROC`).We use three publicly available datasets that have been used in similar studies. Hulth2003 dataset contains abstracts from medical domain, whereas KDD and WWW datasets contain abstracts from computer science domain published in these two well-known conferences. Each document in these datasets is accompanied by an associated gold-standard keywords list, which is used as ground truth for testing the classifier performance. Table 1 briefly describes the datasets along with relevant statistics. For KDD and WWW datasets, we consider only those documents, which contain at least two sentences, and are accompanied by at least one gold-standard keyword. We create the individual training sets from Hulth2003, KDD, and WWW datasets using the methodology described in Section 3.

---

[1]  We set 'percentage' parameter to 300%

[2]  `https://cran.r-project.org/web/packages`

**Table 1.** Overview of the experimental data collections. $|D|$: Number of docs, $L_{avg}$: average doc length, $N_{avg}$: average gold-standard keywords per doc, $K_{avg}$: average percentage of keywords present in the text.

| Collection | $|D|$ | $L_{avg}$ | $N_{avg}$ | $K_{avg}$ | Description |
|---|---|---|---|---|---|
| Hulth2003[3] [9] | 1500 | 129 | 23 | 90.07 | PubMed abstracts from *Inspec* |
| WWW [4] | 1248 | 174 | 9 | 64.97 | CS articles from WWW conference |
| KDD [4] | 704 | 204 | 8 | 68.12 | CS articles from KDD conference |

We designed experiments to:

i.  evaluate the cross-validated performance of keyword classifiers trained on the individual training sets (Section 4.2).
ii. assess predictive power of the trained models over cross-collection and cross-domain datasets (Section 4.2).
iii. evaluate quality of extracted keywords for title construction of scientific papers (Section 4.2).

### 4.2    Results and Discussion

**Evaluating Cross-validated Performance of the Models:** We trained three models on the balanced training sets (Hulth, KDD-B, and WWW-B) and two models on the imbalanced training sets (KDD and WWW) using Naïve Bayes (NB) and Logistic Regression (LR) algorithms (please see Section 3.3 for details). Since a well-written abstract contains the most important facts about scientific research and proxies well for the complete document, we empirically test the system on abstracts from scientific papers. Nevertheless, the system is extendable to full texts. We present 10-fold cross validation results in Table 2, showing precision, recall, and F1-score as performance evaluation metrics. Bold values represent best performance across all models in terms of the 'positive' class[4].

As expected, models trained on balanced training set yield better result as compared to the ones trained on imbalanced set. Thus, we discard the models trained on KDD and WWW training sets from further experiments. Although models trained on WWW-B training set turns out to be the best from cross-validation performance, we also retain models trained on Hulth and KDD-B for assessing the predictive power of the models over unseen documents. In subsequent experiments, we use a naming convention of M-X for all models, where M stands for the model, which is either NB or LR, and X stands

---

for the training set. For example, NB-Hulth is the model trained on Hulth training set using Naïve Bayes classifier.

**Table 2.** Cross-validated classifier performance. NB: Naïve Bayes classifier results, LR: Logistic Regression classifier results, X-B: balanced training set for the corresponding dataset X.

| Training Set | Naïve Bayes (NB) | | | Logistic Regression (LR) | | |
|---|---|---|---|---|---|---|
| | **P** | **R** | **F1** | **P** | **R** | **F1** |
| Hulth | 64.76 | 51.47 | 57.36 | 72.65 | 47.29 | 57.29 |
| KDD | 37.39 | 58.51 | 54.63 | 55.75 | 20.47 | 29.95 |
| WWW | 40.02 | 60.23 | 48.09 | 60.12 | 23.69 | 33.99 |
| KDD-B | 66.93 | 64.55 | 65.72 | 75.43 | 56.80 | 64.80 |
| WWW-B | 69.20 | **66.24** | 67.69 | **76.04** | 61.36 | **67.92** |

**Assessing Cross-collection Predictive Power of the Models:** We test the performance of the six models (three for each NB and LR) on cross-collection test sets from Hulth2003, KDD, and WWW collections. The test sets comprise the unbalanced training sets for Hulth, KDD, and WWW datasets as described in Section 3.2. We apply each model on all three test sets and report their performance in Table 3. For example, the model trained on Hulth training set is tested on all three training sets from Hulth, KDD, and WWW datasets. Table 3 shows results of the individual models on the corresponding test sets. Results are macro-averaged at the dataset level. Bold values indicate best performance in terms of precision, recall, and F1-score for the corresponding test sets.

**Table 3.** Performances of NB and LR models on test sets. P: Precision, R: Recall, and F1: F1-score.

| Models | Hulth2003 | | | KDD | | | WWW | | |
|---|---|---|---|---|---|---|---|---|---|
| | **P** | **R** | **F1** | **P** | **R** | **F1** | **P** | **R** | **F1** |
| NB-Hulth | 66.7 | **58** | 57.5 | 36.6 | 66.4 | 44.7 | 38.5 | 70.4 | 47 |
| NB-KDD-B | 65.9 | 57.1 | 56.3 | 35.7 | 64.3 | 43.3 | 37.7 | 68.7 | 45.8 |
| NB-WWW-B | 66.5 | 55.4 | 55.8 | 36.8 | 63.5 | 43.9 | 38.4 | 67.5 | 46 |
| LR-Hulth | 74 | 52.6 | **57.7** | 41.5 | **66.9** | **48.7** | 43.7 | **70.4** | **51** |
| LR-KDD-B | 75.2 | 45.4 | 52.2 | **45.4** | 57.6 | 47.9 | **47** | 61.5 | 49.9 |
| LR-WWW-B | **75.3** | 47.8 | 54 | 44.9 | 59.1 | 48.1 | 46.6 | 62.9 | 50.3 |

Performance of LR models are relatively better than NB models in terms of F1-score, as reported in Table 3. We observe that models induced by Logistic Regression exhibit better results in terms of precision and models induced by Naïve Bayes exhibit better result in terms of recall. We also observe that the performance of the models are uniform across all datasets irrespective of the training set used. This indicates that the proposed method is independent of the domain or corpora of its training set, and is applicable to any text document. This experiment also establishes that node properties of graph-of-text are effective discriminators to distinguish keywords from non-keywords.

**Recommending Keywords for Title Construction:** We empirically validate our hypothesis that keywords are suitable candidates for generating titles for scientific documents. We experiment with Hulth2003 dataset that contains title and abstract for each document, where titles are clearly distinguishable from the rest of the text. Though KDD and WWW datasets contain title and abstract as well, titles are embedded in a manner that they are not clearly distinguishable. Thus, we present our results using only the Hulth2003 dataset. Moreover, to the best of the authors' knowledge, there is no work in literature that resembles our objective of semi-automatic title construction through automatic keywords extraction. Thus, we provide empirical validation for our experiment and present them below.

To compute the overlapping between keywords and words in title, we first tokenize the title text and remove stopwords from them. Since the proposed KE algorithm uses stemming, we stem the title-words for comparison. We compute macro-averaged precision and recall at a dataset level comparing both these lists of keywords and title-words. We rank the predicted keywords in decreasing order of their probability for the 'positive' class. We use the models trained using Logistic Regression, i.e., LR-Hulth, LR-KDD-B, and LR-WWW-B for our experiment as they clearly outperformed NB models in Table 3. We present our findings in Table 4, where bold values represent best result in terms of precision and recall.

**Table 4.** Overlapping of extracted keywords and title-words using precision and recall. @*k*: extracting top-*k* keywords, @*lenW*: extracting as many keywords as the number of corresponding title-words, P: Precision, R: Recall.

| Models | @5 | | @7 | | @10 | | @*lenW* | |
|---|---|---|---|---|---|---|---|---|
| | **P** | **R** | **P** | **R** | **P** | **R** | **P** | **R** |
| LR-Hulth | **82.12** | **63.37** | **74.41** | **69.51** | 70.61 | **72.07** | **79.24** | **69.22** |
| LR-KDD-B | 76.38 | 56.27 | 71.81 | 60.28 | 70.30 | 61.56 | 75.11 | 58.96 |
| LR-WWW-B | 77.72 | 58.56 | 72.64 | 63.08 | **70.79** | 64.53 | 76.67 | 62.01 |

We present four set of outcomes in Table 4. We extract top-*k* keywords (@*k*) with *k* being 5, 7, and 10 and we extract as many keywords as the number of title-words in the corresponding document (@*lenW*). We kept the number of extracted keywords to a low value, as effective titles tend to be shorter in length [15, 21]. Best precision is obtained when we extract top-5 keywords, and best recall is obtained when we extract top-10 keywords. LR-Hulth model outperforms other two models in all aspects. The results substantiate our claim that keywords are indeed good candidates for title construction.

## 5 Conclusion and Future Work

In this paper, we presented a semi-automatic system to suggest keywords for title generation. Our approach do not generate a title, instead it recommends impactful words for inclusion in the title. We design a supervised framework to automatically extract keywords from single documents. Our KE approach gains from advantages of graph-based keyword extraction techniques, which makes them applicable to texts from any domain or corpora.

The keywords extracted using predictions of the proposed model are then matched against the corresponding title-words from the document. Initial investigation shows a maximum macro-averaged precision of 82% for our dataset when we suggest top-5 extracted keywords, which supports our hypothesis that keywords are indeed good candidates for title construction. Please note that the extracted keywords constitute only nouns and adjectives (Section 3.1). Since title words are not restricted to only nouns and adjectives, our KE approach is expected to miss some words, which explains the 18% loss in precision. This can be improved by including more part-of-speech categories to the text graph after extensively studying the distribution of title-words. As we are reporting our initial investigation in this paper, this part is out of scope and can be considered as a future work.

Top-10 keywords (stemmed) extracted using LR-Hulth model for the abstract of this manuscript are – 'paper', 'system', 'extract', 'titl', 'construct', 'keyword', 'text', 'word', 'document', 'semiautomat'. We constructed the title for the manuscript using these suggested keywords.